**Magnetic Properties of Lithium-Containing Rare Earth Garnets $Li_3RE_3Te_2O_{12}$ (RE = Y, Pr, Nd, Sm-Lu)**

F. Alex Cevallos and R.J. Cava

Department of Chemistry, Princeton University, Princeton NJ 08542

## Abstract

The synthesis, structural description, temperature dependent magnetic susceptibilities and field dependent magnetizations of a series of rare earth garnets of the form $Li_3RE_3Te_2O_{12}$ (RE = Y, La, Pr, Nd, Sm-Lu) are reported. The structure of $Li_3Dy_3Te_2O_{12}$ is refined from powder X-ray diffraction data. The field dependent magnetizations saturate for some of the members by 9 Tesla at 2 K. Of particular interest for further study in this family is the behavior of the Pr, Tb, Dy and Yb garnets.

## Introduction

Geometric magnetic frustration occurs when the crystal structure of a compound inhibits that system's preferred long-range magnetic ordering arrangement. [1] The archetypical example is for antiferromagnetically coupled spins on an equilateral triangular lattice; two spins can easily arrange antiferromagnetically, but the third has no easy choice. This inhibits long-range magnetic ordering to low temperatures, where order does typically emerge, at temperatures lower than the ordering temperature expected based on the Curie-Weiss theta, a measure of the average magnetic coupling strength. As long-range magnetic ordering is one of the most common electronic ground states, frustrated systems where it is inhibited can exhibit a variety of other, unusual ground states such as spin glass, spin ice, and spin liquid. [1, 2, 3, 4, 5]

One structure type that prominently features triangular lattices of magnetic ions is the garnet structure, shown in Figure 1(a). Garnets typically take the form $A_3B_2C_3O_{12}$, where A, B, and C can each be one of a large number of different atoms; the A atoms have 8 oxygen neighbors, and are frequently rare earth elements, the B atoms are frequently (magnetic or nonmagnetic) elements in octahedral coordination with oxygen and the C atoms are typically smaller late main group elements in tetrahedral coordination with oxygen. [6] Depending on the arrangement of magnetic and non-magnetic atoms in the structure, garnets can display a variety of magnetic properties, including ferrimagnetism [7], geometric magnetic frustration [8, 9], spin glass [8], and spin liquid. [2]

A somewhat unusual family of garnets which contain lithium have recently begun to attract attention for their potential application in Li-ion batteries. [10, 11, 12, 13, 14, 15] In these compounds, the lithium atoms occupy a variety of sites with an adjustable number per formula unit, to balance the charge. [16] It is important to note that, by convention, the formulas of Lithium-containing garnets are not written in the typical A-B-C order, but are written C-A-B, with the tetrahedrally-coordinated atom first, and the octahedrally coordinated atom last: e.g. $Li_3Gd_3Te_2O_{12}$, $Li_5Pr_3Sb_2O_{12}$ and $Li_7La_3Zr_2O_{12}$ are all materials with a garnet structure, with the number of Li-atoms adjusted to balance the charge of the octahedrally-coordinated ions, respectively $Te^{6+}$, $Sb^{5+}$ and $Zr^{4+}$. [13, 16, 17] The locations of all the Li atoms in these structures is a matter of some debate, but it is generally agreed that in the case of 3 Li atoms per formula unit, they occupy the standard garnet tetrahedral C-site. As more Li atoms are added to the structure, they begin to occupy nearby non-standard interstitial sites, as shown in Figure 1(c). While the fully-Li-occupied sites are ordered, intermediate Lithium contents result in random site occupancy and thus a degree of structural disorder on all Lithium sites. [10, 16, 17, 18] In the materials of interest here, the Li ions are ordered. [11]

In this work we study the elementary magnetic properties of the $Li_3RE_3Te_2O_{12}$ family of lithium-containing garnets, which has been chosen for study because it displays the garnet phase for nearly the entire rare earth series (Y, Pr, Nd, Sm-Lu) in an ordered structure. To the best of our knowledge, two previous studies have been performed on this subject [19, 20], but both focus on a smaller subset of lanthanides within the lithium garnet family and no full magnetic examination of the entire series has been reported.

**Experimental**

All samples were synthesized by standard solid state reaction. Powders of $Li_2CO_3$ (Alfa Aesar, 99.0%), $TeO_2$ (Alfa Aesar, 99.99%) and the appropriate rare earth oxide (all Alfa Aesar and at least 99.99%) were mixed together in the proper stoichiometric ratio and ground with an agate mortar and pestle for at least 5 minutes, then placed in an alumina crucible. $Li_2CO_3$ was stored at 120°C for 24 hours before use. All rare earth oxides ($Y_2O_3$, $Pr_6O_{11}$, $Nd_2O_3$, $Sm_2O_3$, $Eu_2O_3$, $Gd_2O_3$, $Tb_4O_7$, $Dy_2O_3$, $Ho_2O_3$, $Er_2O_3$, $Tm_2O_3$, $Yb_2O_3$ and $Lu_2O_3$) were heated at 800°C for at least 3 days before use. Samples were placed into a furnace at 800°C and heated in air for two days with intermediate grinding. After one day of heating the temperature of the furnace was slowly increased to 1000°C (with the exception of the Lu compound, which was only heated to 900°C). Samples were removed from the hot furnace and air quenched to room temperature. Samples frequently contained an impurity phase, $Li_4TeO_5$, which was reduced or eliminated with further heating in all cases. Attempts to synthesize the La, Ce and Sc members of this family under our conditions resulted in no detectable garnet phase.

Room-temperature powder X-ray Diffraction (XRD) measurements were taken using a Bruker D8 Advance Eco diffractometer with Cu Kα radiation (λ = 1.5418 Å) and a LynxEye-XE detector. The Bruker EVA program was used for phase identification. Powder LeBail and Rietveld Refinements were performed using Fullprof Suite. Magnetic susceptibility measurements were taken using a Quantum Design Physical Property Measurement System (PPMS) Dynacool with a vibrating sample mount. Samples were placed in plastic sample holders, and measured as polycrystalline powders. Temperature-dependent DC magnetic susceptibility measurements were taken in an applied field of 1,000 Oe between 1.8 and 300 K. Field-dependent susceptibility measurements were taken at 2 K.

## Results and Discussion

### *Structure*

All compounds in the family $Li_3RE_3Te_2O_{12}$ (RE = Y, Pr, Nd, Sm-Lu) were found to crystallize in the space group *Ia-3d* with the garnet structure, in agreement with previous results. [11, 16, 19, 20] The lattice parameters of all compounds as determined by LeBail fits to the room temperature powder diffraction data can be seen in Table 1, and are in good agreement with previous results. Figure 2 shows the relationship between expected ionic radius and lattice parameter, indicating a fairly linear trend and a clear lanthanide contraction, is expected. Some sample LeBail fits are shown in Figure 3. A Rietveld refinement was performed on $Li_3Dy_3Te_2O_{12}$ in order to obtain more in-depth structural information. The Dy compound was selected simply for having what appeared by eye to be the highest-quality powder pattern. Previous studies have suggested that the Li atoms which sit on the tetrahedral interstitial sites can be observed through standard laboratory X-Ray Diffraction. [11] This is somewhat surprising, as the light Li atoms would normally be assumed to be comparatively undetectable in the XRD pattern, and would require neutron diffraction to observe. To test this result, two refinement models were used: One in which all atomic positions were refined, and one in which only the Dysprosium, Oxygen and Tellurium atoms were refined. The resulting fit can be seen in Figure 4, and atomic positions, thermal parameters, and goodness-of-fit factors can be seen in Table 2. These results indicate that the model containing Li atoms is a slightly better fit. Critically for the magnetic properties, the rare earth atoms are fully ordered in one set of crystallographic sites, and the geometry of the magnetic ions is shown in Figure 1(b).

*Magnetism*

The temperature- and field-dependent magnetic susceptibilities were measured for all $Li_3RE_3Te_2O_{12}$ compounds. The temperature-dependent susceptibilities were fit to the Curie-Weiss Law, $\chi = C / (T - \theta_W)$, where $\chi$ is the magnetic susceptibility, $C$ is the Curie constant, and $\theta_W$ is the Weiss temperature. The effective magnetic moment per $RE^{3+}$ ion was then calculated by the equation $\mu_{eff} = \sqrt{8C}$. The resulting values are collected in Table 3 along with (for the purpose of comparison) previously published values derived from low-temperature Curie-Weiss fits of related Gallium and Aluminum garnet phases, when available. As crystal electric field effects may affect the magnetic susceptibility measured at high temperatures, Curie-Weiss fits were performed in both low- and high-temperature regions. The field-dependent magnetic susceptibilities were measured at 2 K and normalized by the moles of $RE^{3+}$ ion present. The magnetic measurements and Curie-Weiss fits for each compound can be seen in Figures 5-15. The magnetic characteristics of the materials studied will now be described on an individual basis.

*$Li_3Pr_3Te_2O_{12}$*

The magnetic susceptibility plots and Curie-Weiss fits are shown in Figure 5. A high-temperature Curie-Weiss fit resulted in values of C = 5.1 and $\mu_{\mathrm{eff}}$ = 3.7 $\mu_B$, in good agreement with the expected value of 3.58 $\mu_B$ for $Pr^{3+}$. The Weiss temperature was found to be -80.3 K, suggesting that $Li_3Pr_3Te_2O_{12}$ has significant antiferromagnetic interactions between spins. A low-temperature Curie-Weiss fit between 10 and 40 K resulted in values of $C$ = 2.8, $\mu_{\mathrm{eff}}$ = 2.7 $\mu_B$ and $\theta_W$ = -10.0 K. No clear signs of ordering are observed down to 1.8 K, indicating that $Li_3Pr_3Te_2O_{12}$ is slightly magnetically frustrated with a frustration index $f = -\theta_W / T_N$ of at least 5, although the flattening of the susceptibility curve at low temperatures hints at a possible transition at lower temperatures. The large difference between the high and low temperature fits is tentatively attributed to CEF effects at higher temperatures. While Applegate et al. report magnetic susceptibility data, they do not report the results of a Curie-Weiss fit to the data. [19] Our field-dependent susceptibility curve measured at 2 K shows a nonlinear response to applied field but no signs of saturation up to 9 Tesla.

*$Li_3Nd_3Te_2O_{12}$*

Temperature- and field-dependent magnetic susceptibility curves for $Li_3Nd_3Te_2O_{12}$ can be seen in Figure 6. A Curie-Weiss fit in the high-temperature regime yielded $C$ = 5.4, $\mu_{eff}$ = 3.7 $\mu_B$ (close to the expected free-ion magnetic moment for $Nd^{3+}$ of 3.62 $\mu_B$) and $\theta_W$ = -68.0 K, indicating of antiferromagnetic coupling between spins. The lower-temperature Curie-Weiss fit yielded values of $C$ = 2.4, $\mu_{eff}$ = 2.5 $\mu_B$ and $\theta_W$ = -1.3 K. These values are consistent with other known Nd-containing frustrated compounds [28, 29, 30] and the difference between high and low temperatures is attributed to low-lying crystal field levels. [31] The low-temperature fit is in good agreement with the previous work by Applegate et al. [19], although the $\theta_W$ determined here via low-temperature fit is somewhat larger than their value of -0.91 K. Our characterization of the magnetization with applied field at 2 K shows a nonlinear response with a degree of saturation of approximately 1.25 $\mu_B$ / $Nd^{3+}$, lower than the expected magnetization $M$ of 3.27 $\mu_B$ but in line with previous observations for $Nd_3Ga_5O_{12}$. [32]

*$Li_3Sm_3Te_2O_{12}$*

A Curie-Weiss fit in the low-temperature region as shown in Figure 7 yielded values of $C$ = 0.14 and $\mu_{eff}$ = 0.61 $\mu_B$. These values are in-line with the expected effective magnetic moment of the free ion $Sm^{3+}$ value of 0.84 $\mu_B$, although somewhat low. This has been observed in other Sm-containing oxides and is attributed to crystal field effects. [31, 33, 34] The fit also yielded a $\theta_W$ = -2.2 K, indicating an antiferromagnetically interacting system, although there are no signs of any magnetic ordering above 1.8 K. A fit at higher temperatures yielded values of $C$ = 4.77, $\mu_{eff}$ = 3.6 $\mu_B$ and $\theta_W$ = -1,412.4 K. Some caution is warranted, however: this implied magnetic moment is more than four times what would be expected for the $Sm^{3+}$ ion, and the curie Weiss temperature is unrealistically large. This kind of behavior has been seen in other Sm-containing systems and is attributed to a non-linear inverse magnetic susceptibility response arising from low-level ionic excited states [35], as well as the temperature-independent Van Vleck paramagnetic contribution which is common in Sm-containing compounds. [36] As a result, this value as well as the large negative Weiss Temperature obtained from the high-temperature fit are not meaningful. The high-temperature values are thus omitted from Table 3. Our characterization of the magnetization with respect to applied field of $Li_3Sm_3Te_2O_{12}$ shows a small degree of curvature, but the magnetization does not saturate at fields up to 9 T.

*$Li_3Eu_3Te_2O_{12}$*

As the $Eu^{3+}$ ion has angular momentum values of L = 3, S = 3 and J = 0, $Eu^{3+}$ has an expected magnetic moment of 0 $\mu_B$. However, the magnetic susceptibility appears to show some temperature dependence as seen in Figure 8. This is due to a combination of factors; the region between approximately 10 and 150 K is dominated by temperature-independent Van Vleck paramagnetism arising from the non-magnetic ground state. Above this temperature range, crystal field states have a visible effect and there is a clear temperature dependence. This is typical in $Eu^{3+}$ systems [31, 33] and while the resulting curve appears similar in shape to compounds with Curie-Weiss behavior, it is unrelated and a fit to the Curie-Weiss law will not result in any meaningful information. The increase in the magnetic susceptibility at very low temperatures is a common feature of $Eu^{3+}$ systems and is typically attributed to a combination of subtle magnetic interactions and small amounts of otherwise undetectable magnetic impurity. [31, 33, 37] The magnetization with respect to temperature shows a very slight curvature at lower fields, but is almost entirely linear up to applied fields of 9 Tesla.

*$Li_3Gd_3Te_2O_{12}$*

The magnetic susceptibility measurements for $Li_3Gd_3Te_2O_{12}$ can be seen in Figure 9. The inverse susceptibility with respect to temperature was well fit to the Curie-Weiss Law with no temperature independent term. In the high-temperature regime, the effective magnetic moment was determined to be 7.8 $\mu_B$, and at low temperatures it was found to be 7.9 $\mu_B$, both values in good agreement with the expect $\mu_{eff}$ for $Gd^{3+}$ of 7.94 $\mu_B$. The Curie constant $C$ was found to be 22.9 in the high-temperature region, and 23.2 at low temperatures. The Weiss temperature $\theta_W$ was found to be -2.0 K in the high-temperature region and -2.9 K in the low-temperature region, both indicating antiferromagnetic interactions. These low-temperature fit values are in good agreement with both Applegate and Mukherjee [19, 20], although Mukherjee et al. find a somewhat high effective moment of 8.233 $\mu_B$. In addition, Applegate et al. observe an ordering transition at T = 0.25 K. The magnetization vs applied field curve begins to saturate at fields of approximately 3 Tesla, with a final saturation value of approximately 6.5 $\mu_B$ / $Gd^{3+}$. This is similar to what was observed by Mukherjee et al. [20], and slightly lower than the expected saturation magnetization of 7 $\mu_B$.

*$Li_3Tb_3Te_2O_{12}$*

A high-temperature Curie-Weiss fit was performed for $Li_3Tb_3Te_2O_{12}$ with no temperature independent term, resulting in values of C = 36.5, $\theta_W$ = -9.8 K, and $\mu_{eff}$ = 9.8 $\mu_B$. This value for effective magnetic moment is in good agreement with the expected value for the free $Tb^{3+}$ ion of 9.75 $\mu_B$. The negative Weiss Temperature is indicative of dominantly antiferromagnetic coupling between spins, but the lack of clear ordering above 1.8 K suggests that $Li_3Tb_3Te_2O_{12}$ may be slightly magnetically frustrated with a frustration index $f$ of over 5. A heat capacity measurement conducted by Mukherjee et al. suggests magnetic ordering at 1.04 K, [20] which would indicate a frustration factor of around 9. At low temperatures, a Curie-Weiss fit yields values of $C$ = 30.6, $\mu_{eff}$ = 9.0 $\mu_B$, and $\theta_W$ = -0.30 K. The values obtained in the low-temperature fit correspond well to the results obtained by both Applegate and Mukherjee [19, 20], although $\theta_W$ is much larger in magnitude than the value from Applegate et al. of -0.017 K. The magnetization vs applied field curve in Figure 10 shows some saturation at high fields, at approximately 5.5 $\mu_B$ / $Tb^{3+}$ ion, which is much lower than the expected saturation magnetization of 9 $\mu_B$, although more in line with the values recorded for $Tb_3Ga_5O_{12}$. [38] Two small anomalies can be seen in in the curve- one at approximately 1 Tesla, and the other at approximately 3 Tesla. The M vs H curve shows a small degree of hysteresis. These anomalies can be clearly seen in the data obtained by Mukherjee et al. [20], but they do not report an M-H "loop" measurement.

*$Li_3Dy_3Te_2O_{12}$*

The DC magnetic susceptibility curves of $Li_3Dy_3Te_2O_{12}$ can be seen in Figure 11. In an applied field of 1,000 Oe, a high-temperature Curie-Weiss fit yields values of C = 42.4, $\mu_{eff}$ = 10.6 $\mu_B$ (in good agreement with the expected free ion value of 10.63 $\mu_B$) and $\theta_W$ = 10.7 K. A low-temperature fit yields similar values of C = 38.3, $\mu_{eff}$ = 10.1 $\mu_B$, and a negative $\theta_W$ of -3.6 K. The low-temperature results are in good agreement with those of Mukherjee et al. [20] (although they obtain a smaller $\theta_W$ of -1.52 K), and somewhat greater in magnitude than those of Applegate et al. ($C$ = 33.72 per formula unit, $\theta_W$ = -0.78 K). [19] A clear antiferromagnetic ordering transition can be observed at ~2 K, shown in the inset of Figure 11. This transition is suppressed by higher magnetic fields, and is not observed in an applied field of 10,000 Oe. $Li_3Dy_3Te_2O_{12}$ is the only compound described here with a visible magnetic ordering temperature above 1.8 K. No other compounds displayed a clear transition, even in applied fields as low as 20 Oe. Interestingly, this transition was not observed by Mukherjee in the magnetic susceptibility data, but is seen in the heat capacity data where there is a clear ordering transition at 1.97 K [20].

Applegate, on the other hand, clearly observes the transition in the magnetic susceptibility data [19]. The magnetization vs applied field curve of $Li_3Dy_3Te_2O_{12}$ is highly non-linear, with an onset of saturation at approximately 1 Tesla, and a magnetization of approximately 6 $\mu_B$ / $Dy^{3+}$ (and still visibly increasing) in an applied field of 9 T, similar to what was observed by Mukherjee et al. [20] although much lower than the expected value of 10 $\mu_B$. A similar low value for saturation has been observed in the related system $Dy_3Ga_5O_{12}$. [39]

*$Li_3Ho_3Te_2O_{12}$*

The temperature- and field-dependent magnetic susceptibility curves for $Li_3Ho_3Te_2O_{12}$ are shown in Figure 12. A Curie-Weiss fit in the high-temperature regime yields values of $C$ = 39.1, $\mu_{eff}$ = 10.2 $\mu_B$, and $\theta_W$ = 9.0 K. The effective magnetic moment of 10.2 $\mu_B$ is in agreement with the expected value for a free $Ho^{3+}$ ion of 10.60 $\mu_B$. A low-temperature fit yields values of $C$ = 31.9, $\mu_{eff}$ = 9.2 $\mu_B$, and $\theta_W$ = -1.6 K. The negative Weiss temperature suggests antiferromagnetic coupling at very low temperatures. These low-temperature values are somewhat greater in magnitude than those of both Applegate and Mukherjee. [19, 20] Mukherjee also observes a possible magnetic ordering transition in the heat capacity data at 1.4 K, which is consistent with our measured Weiss temperature and would not indicate any notable degree of magnetic frustration. [20] The magnetization vs. applied field curve shows a strongly non-linear response, with saturation visible above fields of 1 Tesla (although still increasing up to 9 T), and a magnetization of about 5.5 $\mu_B$ / $Ho^{3+}$ at 9 T, much lower than the expected value of 10 $\mu_B$, but in line with the results of Mukherjee et al. [20] Both results are lower than the (still lower than predicted) saturation value of 7.69 $\mu_B$ observed in $Ho_3Ga_5O_{12}$. [38]

*$Li_3Er_3Te_2O_{12}$*

The magnetic susceptibility data for $Li_3Er_3Te_2O_{12}$ is shown in Figure 13. A high-temperature Curie-Weiss fit yielded values of $C$ = 31.0, $\theta_W$ = 2.0 K, and $\mu_{eff}$ = 9.1 $\mu_B$, in good agreement with the expected value of 9.59 $\mu_B$ for the $Er^{3+}$ ion. A Curie-Weiss fit at lower temperatures yielded values of $C$ = 20.2, $\mu_{eff}$ = 7.3 $\mu_B$ and $\theta_W$ = -0.42 K. The low-temperature value for $C$ is slightly larger than the corresponding formula unit value of $C$ from Applegate et al. (approximately 18.5), and $\theta_W$ is fairly close to their obtained value of -0.28 K. [19] The magnetization with respect to applied field is highly non-linear, with saturation visible at fields above 1.5 Tesla, and reaching a magnitude of approximately 4.5 $\mu_B$ at applied fields of 9 Tesla, much lower than the expected saturation value of 9 $\mu_B$, although it is noted that in the related compound $Er_3Ga_5O_{12}$, full magnetic saturation is not observed even in applied fields of over 30 T. [40]

*$Li_3Tm_3Te_2O_{12}$*

Figure 14 shows the magnetization data for $Li_3Tm_3Te_2O_{12}$. In the high-temperature regime, a Curie-Weiss yields values of $C$ = 19.6, $\mu_{eff}$ = 7.2 $\mu_B$ and $\theta_W$ = 5.9 K. The measured effective magnetic moment is fairly close to the expected value of 7.57 $\mu_B$ for the free $Tm^{3+}$ ion. A Curie-Weiss fit in the low-temperature regime (60-100 K) yields values of $C$ = 24.9, $\mu_{eff}$ = 8.1 $\mu_B$ and $\theta_W$ = -38.3 K. The negative Weiss temperature suggests antiferromagnetic ordering, and as there is no clear magnetic ordering transition above 1.8 K, this suggests magnetic frustration with a frustration index $f$ of 21 or greater, although at these temperatures this may still result from CEF effects. A plateau-like feature is visible in the temperature-dependent susceptibility data below 30 K. This type of feature is not unusual in Tm-containing compounds [33], and can be interpreted as Van Vleck contribution to the magnetic susceptibility arising from a crystal field splitting-induced singlet ground state of the $Tm^{3+}$ ion. [41] The field-dependent magnetization data is highly linear with no visible saturation.

*$Li_3Yb_3Te_2O_{12}$*

Susceptibility vs. temperature and magnetization vs. applied field curves for $Li_3Yb_3Te_2O_{12}$ can be seen in Figure 15. A Curie-Weiss fit in the high-temperature regime yielded $C$ = 7.3, $\mu_{\text{eff}}$ = 4.4 $\mu_B$ (close to the expected free ion value of 4.54 $\mu_B$) and $\theta_W$ = -56.4 K. This large negative Weiss temperature is indicative of antiferromagnetic coupling, and the lack of an ordering transition suggests that the compound may be magnetically frustrated, with a frustration index $f$ of over 30. At low temperatures, the Curie-Weiss fit yields lower values of $C$ = 3.7, $\mu_{\text{eff}}$ = 3.1 $\mu_B$, and a $\theta_W$ of only -0.43 K. These low-temperature fit values are in line with those obtained by Applegate et al. [19], although the value of $\theta_W$ is much higher than their value of -0.062 K. This is consistent with previous observations of the $Yb^{3+}$ ion low-temperature Kramers doublet ground state. [42, 43] This ground state is indicative of magnetic anisotropy, which has long been known to occur in other Yb-containing garnets such as $Yb_3Fe_5O_{12}$. [44] The magnetization with respect to applied field is highly non-linear and shows clear saturation in fields above 2 Tesla, with a saturation moment of approximately 1.75 $\mu_B$. This is much lower than the expected saturation value of 4 $\mu_B$, but corresponds well to previous measurements on $Yb_3Ga_5O_{12}$. [45]

*Discussion*

The measured effective magnetic moments of the $Li_3RE_3Te_2O_{12}$ family, both high- and low-temperature values, are plotted in Figure 16, along with the ideal values for free ion magnetic moments. In the garnet structure, the A-site (occupied by the RE elements in this case) is 8-fold coordinated; in an effort to observe any effects of coordination geometry and crystal electric field on the resulting magnetic moment, this data is plotted alongside the available values of magnetic moments for an 8-fold coordinated RE compound ($RE_2Ti_2O_7$ pyrochlore) and a 6-fold, octahedrally-coordinated RE compound ($KBaRE(BO_3)_2$ in the low-temperature case, $Ba_2RE_{2/3}TeO_6$ in the high-temperature case). The plotted values can be seen in Table 4. It is observed that at high temperatures, the magnetic moments of all compounds are approximately equivalent, and close in value to the ideal free ion magnetic moment.

The story is quite different in the low-temperature data; immediately apparent is the fact that the measured moments at low temperatures are consistently lower than the ideal free ion values, often by significant margins. The singular and quite notable exception to this trend is Tm, where the measured values for both the 8-fold garnet and 6-fold coordinated compounds are *higher* than the ideal value (data is unavailable for the 8-fold coordinated pyrochlore $Tm_2Ti_2O_7$). In addition to this anomaly, the values for the three structure families vary much more than in the high-temperature case, with sometimes large separations between values for the same element. No unambiguously clear trends are immediately apparent, although it is noted that the effective moments in the 6-fold coordinated borates tend to be relatively close to the ideal values, while the 8-fold coordinated garnets show a large degree of element-dependent variability. Finally, it is noted that the Er-containing compounds seem to show the most variation with change in structure, with the 8-fold coordinated pyrochlore having almost the ideal magnetic moment, while the 8-fold coordinated garnet is over 20% lower than the ideal value.

The magnetization at 9 T in cases where saturation or near saturation was observed in the M-H curves are plotted in Figure 17, along with the "ideal" values of saturation magnetization, $M = g_J*J$. The lower panel shows the same data normalized to the "ideal value". It can be seen that the Gd case is closest to the expected saturation value but that many of the materials display magnetizations that are considerably less than what is expected. As in other materials in which similar behavior has been observed [52], in the current materials it may be due to the presence of magnetic anisotropy and thus a powder averaging over crystallites with different orientations with respect to the applied field. This is consistent with the observation that the $4f^7$ Gd-based material, expected to have the most isotropic spin system, shows saturation closest to the ideal full value.

**Conclusion**

A series of rare-earth containing lithium garnets of the form $Li_3Ln_3Te_2O_{12}$ (*Ln* = Y, Pr, Nd, Sm-Lu) has been synthesized. The lattice parameter *a* has been determined for each compound via LeBail fit on room-temperature Powder X-Ray data, and the results, where they overlap, are in good agreement with previously published values and demonstrate a clear lanthanide contraction. Magnetic susceptibility measurements have been taken, and Curie-Weiss fits performed. Several of the compounds observed exhibit some degree of magnetic frustration, and only $Li_3Dy_3Te_2O_{12}$ shows a clear magnetic ordering transition above 2 K. Many of our results are in reasonable agreement with previous studies [19, 20], but several significant disparities have been noted. As lithium-containing garnets continue to show promise as Li-ion conductors, further magnetic characterization of these compounds, particularly in the form of single crystals, may provide valuable insights into their behavior and possibly point to additional applications beyond battery materials.

**Acknowledgements**

This research was performed under the auspices of the Institute for Quantum Matter and supported by the US Department of Energy, Division of Basic Energy Sciences, Grant No. DE-FG02-08ER46544.

## Tables

**Table 1.** Lattice parameters of $Li_3RE_3Te_2O_{12}$ compounds, as determined by LeBail fits to powder X-Ray Diffraction data at ambient temperature, compared with the effective ionic radii of the rare earth elements. All compounds crystallize in the space group *Ia-3d* (no. 230).

| **Rare Earth element** | **RE ionic radius (Å)** | **Lattice parameter *a* (Å)** |
|---|---|---|
| Y | .900 | 12.269(1) |
| Pr | .990 | 12.609(1) |
| Nd | .983 | 12.551(1) |
| Sm | .958 | 12.462(1) |
| Eu | .947 | 12.426(1) |
| Gd | .935 | 12.385(1) |
| Tb | .923 | 12.342(1) |
| Dy | .912 | 12.309(1) |
| Ho | .901 | 12.269(1) |
| Er | .890 | 12.232(1) |
| Tm | .880 | 12.209(1) |
| Yb | .868 | 12.171(1) |
| Lu | .861 | 12.151(1) |

**Table 2.** Lattice parameter, atomic positions, thermal parameters, and goodness-of-fit information for Rietveld refinement of $Li_3Dy_3Te_2O_{12}$ at ambient temperature, comparing two models: In Model 1, all four atomic positions are refined. In Model 2, only Dy, Te and O positions are refined, and Li is left out of the refinement entirely.

| | **Model 1** | | | | **Model 2** | | | |
|---|---|---|---|---|---|---|---|---|
| **Lattice parameter *a* (Å)** | 12.305(1) | | | | 12.305(1) | | | |
| **Atom** | ***x*** | ***y*** | ***z*** | **$B_{iso}$** | ***x*** | ***y*** | ***z*** | **$B_{iso}$** |
| Dy | .125 | 0 | .25 | 2.43(2) | .125 | 0 | 0 | 2.34(2) |
| Te | 0 | 0 | 0 | 2.30(2) | 0 | 0 | 0 | 2.17(2) |
| O | .02623(27) | .05101(30) | .64565(28) | 2.28(11) | .02817(28) | .04988(32) | .64757(29) | 2.75(12) |
| Li | .375 | 0 | .25 | 4.19(53) | - | - | - | - |
| **Rietveld R-factor** | | | | | | | | |
| $R_f$ | 8.29 | | | | 7.73 | | | |
| $R_p$ | 16.7 | | | | 17.8 | | | |
| $R_{wp}$ | 15.5 | | | | 16.0 | | | |
| $\chi^2$ | 3.25 | | | | 3.46 | | | |

**Table 3**. Values determined by Curie-Weiss fits for $Li_3RE_3Te_2O_{12}$ compounds in an applied field of 1,000 Oe at high temperature (150-300 K) and low temperature (10-40 K). Y, Eu, and Lu compounds excluded. Values of $\mu_{eff}$ and $\theta_W$ from low-temperature fits for RE-containing Gallium and Aluminum garnets are provided for comparison, where available.

| Compound | Temperature Range | Curie Constant | $\theta_W$ (K) | $\mu_{eff}$ ($\mu_B/RE^{3+}$) |
|---|---|---|---|---|
| $Li_3Pr_3Te_2O_{12}$ | High | 5.1 | -80.3 | 3.7 |
| | Low | 2.8 | -10.0 | 2.7 |
| *$Pr_3Ga_5O_{12}$* [a] | | | - | *2.88* |
| $Li_3Nd_3Te_2O_{12}$ | High | 5.3 | -68.0 | 3.7 |
| | Low | 2.4 | -1.3 | 2.5 |
| *$Nd_3Ga_5O_{12}$* [a] | | | - | *2.71* |
| $Li_3Sm_3Te_2O_{12}$ | High | - | - | - |
| | Low | 0.14 | -2.2 | 0.61 |
| *$Sm_3Ga_5O_{12}$* [b] | | | *-2.5* | *0.5* |
| $Li_3Gd_3Te_2O_{12}$ | High | 22.9 | -2.0 | 7.8 |
| | Low | 23.2 | -2.9 | 7.9 |
| *$Gd_3Ga_5O_{12}$* [c] | | | *-1.4* | *7.814* |
| *$Gd_3Al_5O_{12}$* [d] | | | *-3.0* | *7.91* |
| $Li_3Tb_3Te_2O_{12}$ | High | 36.5 | -9.8 | 9.8 |
| | Low | 30.6 | -0.30 | 9.0 |
| *$Tb_3Ga_5O_{12}$* [c] | | | *-1.16* | *8.34* |
| $Li_3Dy_3Te_2O_{12}$ | High | 42.4 | 10.7 | 10.6 |
| | Low | 38.3 | -3.6 | 10.1 |
| *$Dy_3Ga_5O_{12}$* [c] | | | *-1.00* | *8.34* |
| *$Dy_3Al_5O_{12}$* [e] | | | *0.20* | *9.1* |
| $Li_3Ho_3Te_2O_{12}$ | High | 39.1 | 9.0 | 10.2 |
| | Low | 31.9 | -1.6 | 9.2 |
| *$Ho_3Ga_5O_{12}$* [c] | | | *-3.29* | *10.3* |
| $Li_3Er_3Te_2O_{12}$ | High | 31.0 | 2.0 | 9.1 |
| | Low | 20.2 | -0.42 | 7.3 |
| *$Er_3Ga_5O_{12}$* [f] | | | - | *8.4* |
| $Li_3Tm_3Te_2O_{12}$ | High | 19.6 | 5.9 | 7.2 |
| | Low | 24.9 | -38.3 | 8.1 |
| *$Tm_3Ga_5O_{12}$* [g] | | | - | *7.5* |
| $Li_3Yb_3Te_2O_{12}$ | High | 7.3 | -56.4 | 4.4 |
| | Low | 3.7 | -0.43 | 3.1 |
| *$Yb_3Ga_5O_{12}$* [h] | | | *-0.048* | *2.99* |
| *$Yb_3Al_5O_{12}$* [h] | | | *-0.139* | *2.98* |

a. ref [21]
b. ref [22]
c. ref [20]
d. ref [23]
e. ref [24]
f. ref [25]
g. ref [26]
h. ref [27]

**Table 4.** Effective magnetic moments for several families of RE-containing compounds at high and low temperatures, as well as the ideal values for free ion magnetic moments. All values in $\mu_B$.

| | | High-T | | | Low-T | | |
|---|---|---|---|---|---|---|---|
| **RE element** | **Ideal** | **$Li_3RE_3Te_2O_{12}$** | **$RE_2Ti_2O_7$** | **$Ba_2RE_{2/3}TeO_6$** | **$Li_3RE_3Te_2O_{12}$** | **$RE_2Ti_2O_7$** | **$KBaRE(BO_3)_2$** |
| Pr | 3.58 | 3.7 | - | - | 2.7 | - | - |
| Nd | 3.62 | 3.7 | - | 3.3[e] | 2.5 | - | - |
| Sm | 0.84 | - | - | - | 0.61 | - | 0.43[g] |
| Gd | 7.94 | 7.8 | 7.93[a] | 7.8[e] | 7.9 | 7.22[f] | 7.7[g] |
| Tb | 9.75 | 9.8 | 9.68[a] | 9.5[e] | 9 | - | 9.68[g] |
| Dy | 10.63 | 10.6 | 10.51[b] | 10.4[e] | 10.1 | 9.615[f] | 10.07[g] |
| Ho | 10.6 | 10.2 | 10.48[c] | 10.5[e] | 9.2 | 9.27[f] | 10.09[g] |
| Er | 9.59 | 9.1 | 9.47[d] | 9.5[e] | 7.3 | 9.34[f] | 8.86[g] |
| Tm | 7.57 | 7.2 | - | 7.6[e] | 8.1 | - | 7.8[g] |
| Yb | 4.54 | 4.4 | - | 4.7[e] | 3.1 | 3.335[f] | 2.67[g] |

a. ref [46]

b. ref [47]

c. ref [48]

d. ref [49]

e. ref [50]

f. ref [51]

g. ref [33]

**Figure Captions**

**Figure 1.** (a) The unit cell of the garnet $Li_3RE_3Te_2O_{12}$. Li-O tetrahedra are green, Te-O octahedra are brown, $RE^{3+}$ ions are blue. (b) A view of only the $RE^{3+}$ ions in the crystal structure, showing that the RE lattice is composed of interpenetrating rings of corner-sharing triangles. The RE ions are all crystallographically equivalent; they have been colored to illustrate the interpenetrating ring structure. (c) Comparison of Li-atom positions in a portion of the unit cell of $Li_3RE_3Te_2O_{12}$ (left) and $Li_7RE_3Zr_2O_{12}$ (right, Zr-O octahedra shown in orange). [18]

**Figure 2.** The lattice parameter *a* as a function of rare earth ionic radius for $Li_3RE_3Te_2O_{12}$.

**Figure 3.** Example LeBail fits to room-temperature Powder XRD measurements for two $Li_3RE_3Te_2O_{12}$ compounds. The measured data is in red, the calculated patterns are black, the Bragg reflections are green, and the difference plots are blue. Left: $Li_3Nd_3Te_2O_{12}$. Right: $Li_3Ho_3Te_2O_{12}$. Cu Kα radiation.

**Figure 4.** Rietveld refinement of $Li_3Dy_3Te_2O_{12}$ Powder XRD data at room temperature. Cu Kα radiation.

**Figure 5.** Left: Magnetic susceptibility and inverse susceptibility of $Li_3Pr_3Te_2O_{12}$ with respect to temperature in an applied field of 1,000 Oe. Curie-Weiss fits shown in black. Right: Field-dependent magnetization of $Li_3Pr_3Te_2O_{12}$ measure at 2 K. Inset: Low-temperature region of inverse magnetic susceptibility with respect to temperature.

**Figure 6.** Left: Magnetic susceptibility and inverse susceptibility of $Li_3Nd_3Te_2O_{12}$ with respect to temperature in an applied field of 1,000 Oe. Curie-Weiss fits shown in white. Right: Field-dependent magnetization of $Li_3Nd_3Te_2O_{12}$ measure at 2 K. Inset: Low-temperature region of inverse magnetic susceptibility with respect to temperature.

**Figure 7.** Left: Magnetic susceptibility and inverse susceptibility of $Li_3Sm_3Te_2O_{12}$ with respect to temperature in an applied field of 1,000 Oe. Curie-Weiss fits shown in black. Right: Field-dependent magnetization of $Li_3Sm_3Te_2O_{12}$ measure at 2 K. Inset: Low-temperature region of inverse magnetic susceptibility with respect to temperature.

**Figure 8.** Left: Magnetic susceptibility and inverse susceptibility of $Li_3Eu_3Te_2O_{12}$ with respect to temperature in an applied field of 1,000 Oe. Right: Field-dependent magnetization of $Li_3Eu_3Te_2O_{12}$ measure at 2 K. Inset: Low-temperature region of inverse magnetic susceptibility with respect to temperature.

**Figure 9.** Left: Magnetic susceptibility and inverse susceptibility of $Li_3Gd_3Te_2O_{12}$ with respect to temperature in an applied field of 1,000 Oe. Curie-Weiss fits shown in white. Right: Field-dependent magnetization of $Li_3Gd_3Te_2O_{12}$ measure at 2 K. Inset: Low-temperature region of inverse magnetic susceptibility with respect to temperature.

**Figure 10.** Left: Magnetic susceptibility and inverse susceptibility of $Li_3Tb_3Te_2O_{12}$ with respect to temperature in an applied field of 1,000 Oe. Curie-Weiss fits shown in white. Right: Field-dependent magnetization of $Li_3Tb_3Te_2O_{12}$ measure at 2 K. Inset: Low-temperature region of inverse magnetic susceptibility with respect to temperature.

**Figure 11.** Left: Magnetic susceptibility and inverse susceptibility of $Li_3Dy_3Te_2O_{12}$ with respect to temperature in an applied field of 1,000 Oe. Curie-Weiss fits shown in white. Right: Field-dependent magnetization of $Li_3Dy_3Te_2O_{12}$ measure at 2 K. Inset: Low-temperature region of magnetic susceptibility and inverse magnetic susceptibility with respect to temperature.

**Figure 12.** Left: Magnetic susceptibility and inverse susceptibility of $Li_3Ho_3Te_2O_{12}$ with respect to temperature in an applied field of 1,000 Oe. Curie-Weiss fits shown in white. Right: Field-

dependent magnetization of $Li_3Ho_3Te_2O_{12}$ measure at 2 K. Inset: Low-temperature region of inverse magnetic susceptibility with respect to temperature.

**Figure 13.** Left: Magnetic susceptibility and inverse susceptibility of $Li_3Er_3Te_2O_{12}$ with respect to temperature in an applied field of 1,000 Oe. Curie-Weiss fits shown in black. Right: Field-dependent magnetization of $Li_3Er_3Te_2O_{12}$ measure at 2 K. Inset: Low-temperature region of inverse magnetic susceptibility with respect to temperature.

**Figure 14.** Left: Magnetic susceptibility and inverse susceptibility of $Li_3Tm_3Te_2O_{12}$ with respect to temperature in an applied field of 1,000 Oe. Curie-Weiss fits shown in white. Right: Field-dependent magnetization of $Li_3Tm_3Te_2O_{12}$ measure at 2 K. Inset: Low-temperature region of inverse magnetic susceptibility with respect to temperature.

**Figure 15.** Left: Magnetic susceptibility and inverse susceptibility of $Li_3Yb_3Te_2O_{12}$ with respect to temperature in an applied field of 1,000 Oe. Curie-Weiss fits shown in white. Right: Field-dependent magnetization of $Li_3Yb_3Te_2O_{12}$ measure at 2 K. Inset: Low-temperature region of inverse magnetic susceptibility with respect to temperature.

**Figure 16.** (a) High-temperature magnetic moments of $Li_3RE_3Te_2O_{12}$ (red), $RE_2Ti_2O_7$ (blue), and $Ba_2RE_{2/3}TeO_6$ (dark yellow) compared to the ideal values of free ion magnetic moment (empty circles). (b) Low-temperature magnetic moments of $Li_3RE_3Te_2O_{12}$ (red), $RE_2Ti_2O_7$ (blue) and $KBaRE(BO_3)_2$ (dark yellow) compared to the ideal values of free ion magnetic moment (empty circles).

**Figure 17**. (a) Measured values of magnetization in an applied field of 9 T in $Li_3RE_3Te_2O_{12}$, for each RE element where saturation or near saturation was observed in the M-H curve at 2 K. The ideal values of saturation magnetization for each element, calculated by $g_J$*J, are plotted for comparison. (b) The measured values of magnetization as fractions of the ideal saturation values. Lines corresponding to fractions of 1/3, 1/2 and 2/3 magnetization have been added for reference.

**Figures**

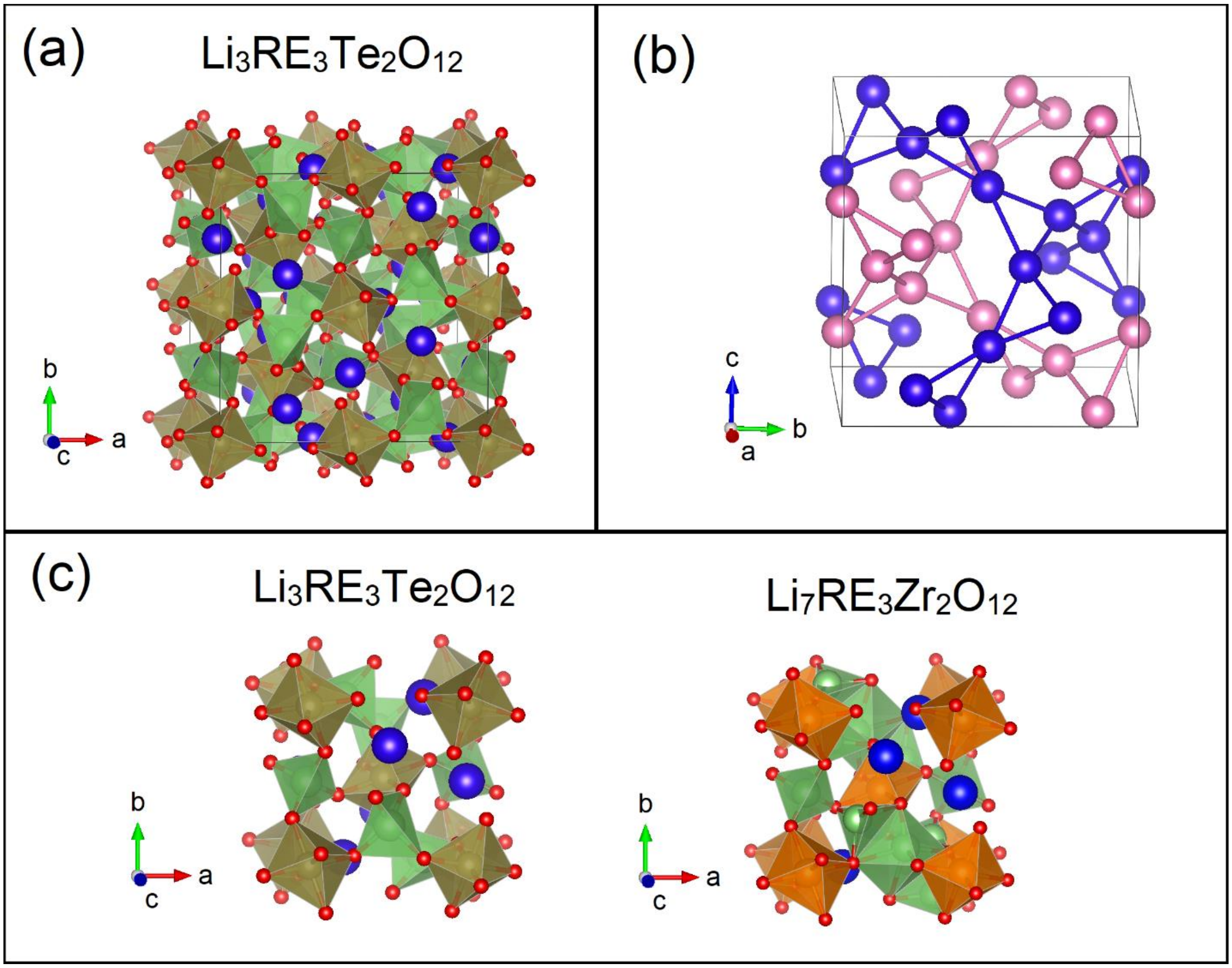

**Figure 1.**

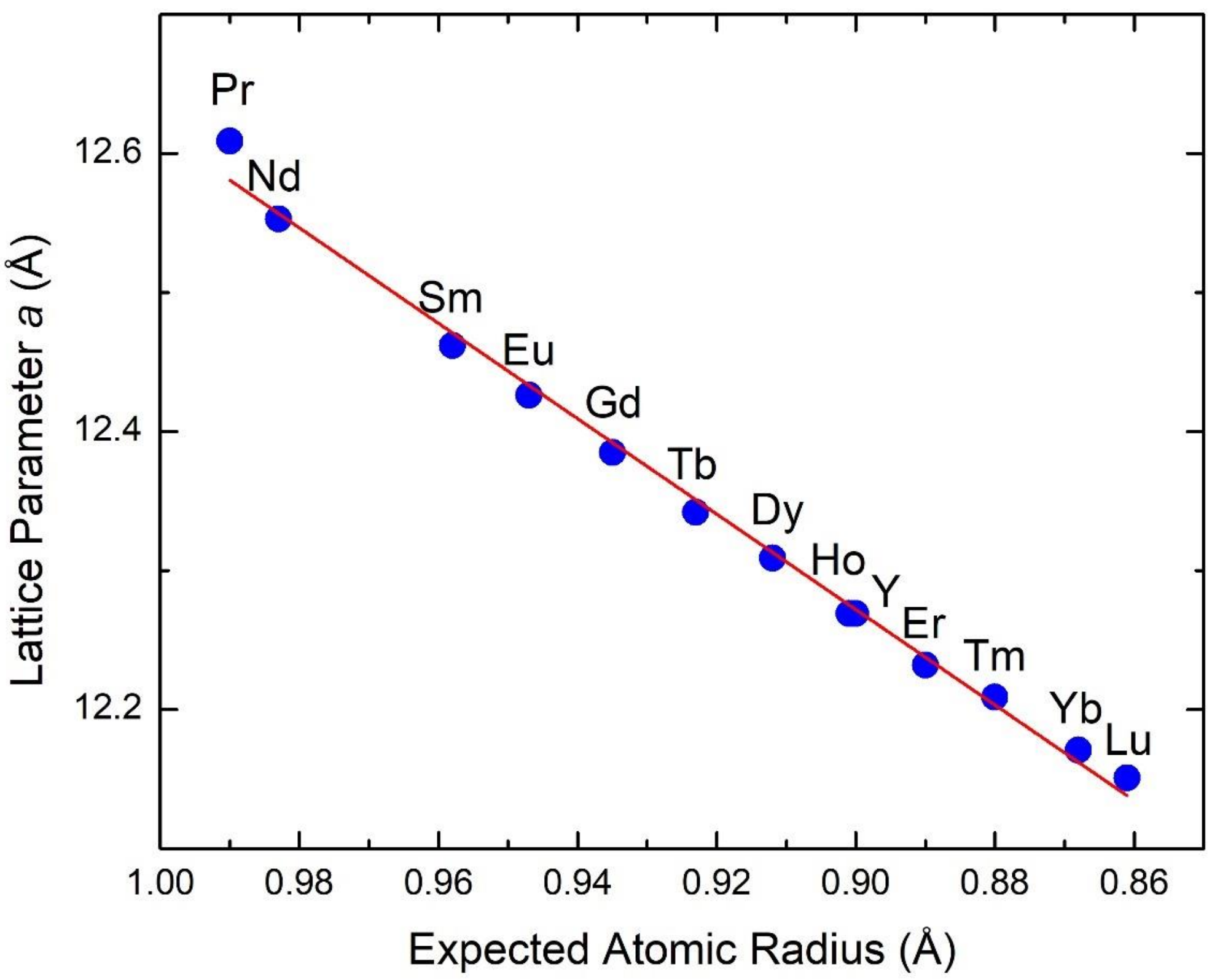

**Figure 2.**

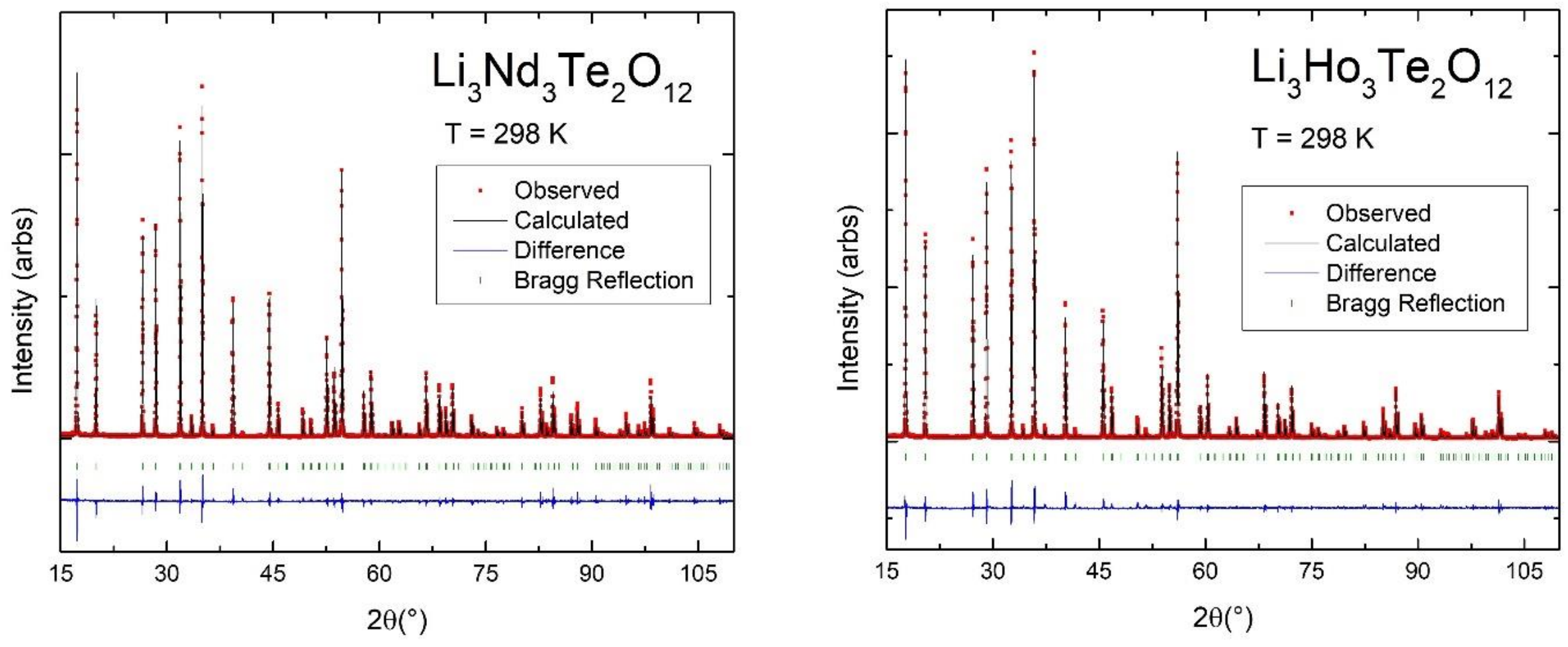
Li3Nd3Te2O12
T = 298 K
Observed
Calculated
Difference
Bragg Reflection
Intensity (arbs)
2θ(°)
15 30 45 60 75 90 105
Li3Ho3Te2O12
T = 298 K
Observed
Calculated
Difference
Bragg Reflection
Intensity (arbs)
2θ(°)
15 30 45 60 75 90 105

**Figure 3.**

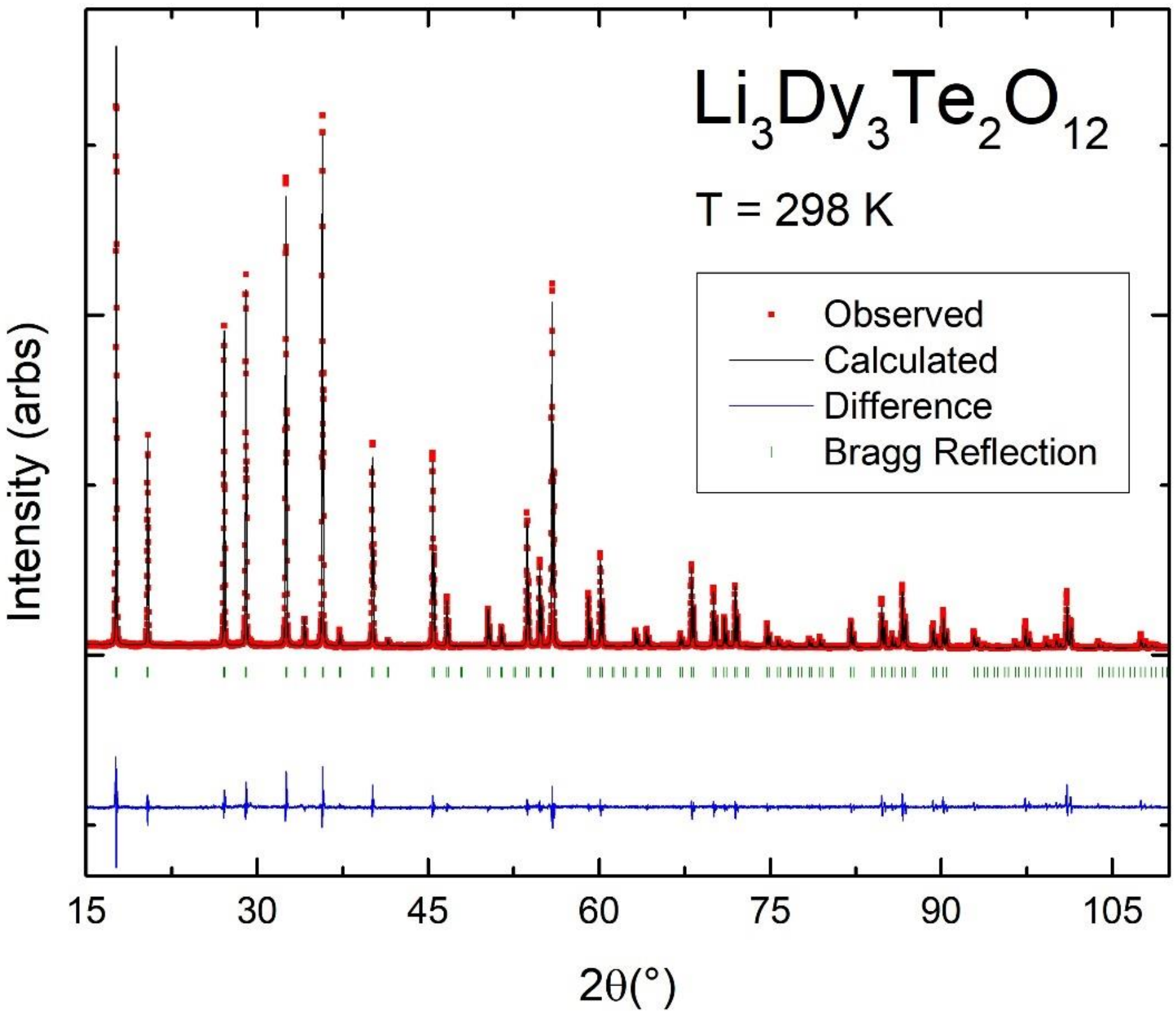

**Figure 4**.

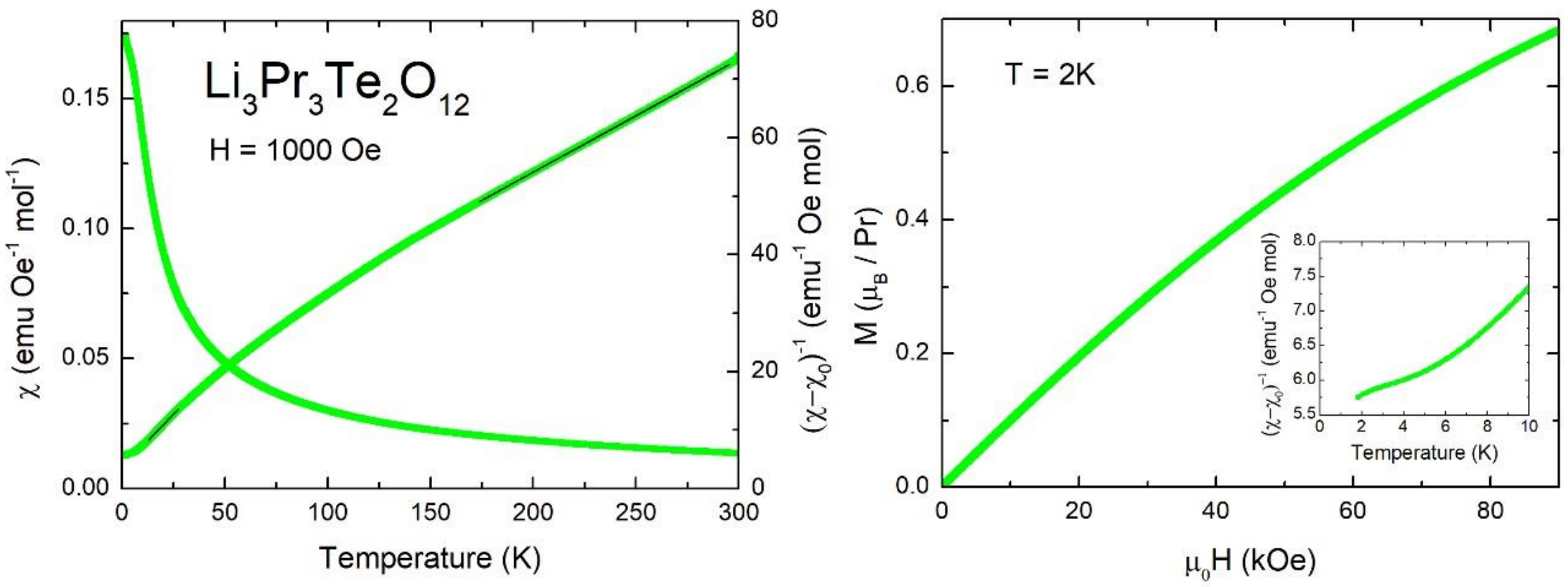

Li3Pr3Te2O12
H = 1000 Oe
χ (emu Oe-1 mol-1)
(χ-χ0)-1 (emu-1 Oe mol)
Temperature (K)
T = 2K
M (μB / Pr)
μ0H (kOe)
(χ-χ0)-1 (emu-1 Oe mol)
Temperature (K)

**Figure 5.**

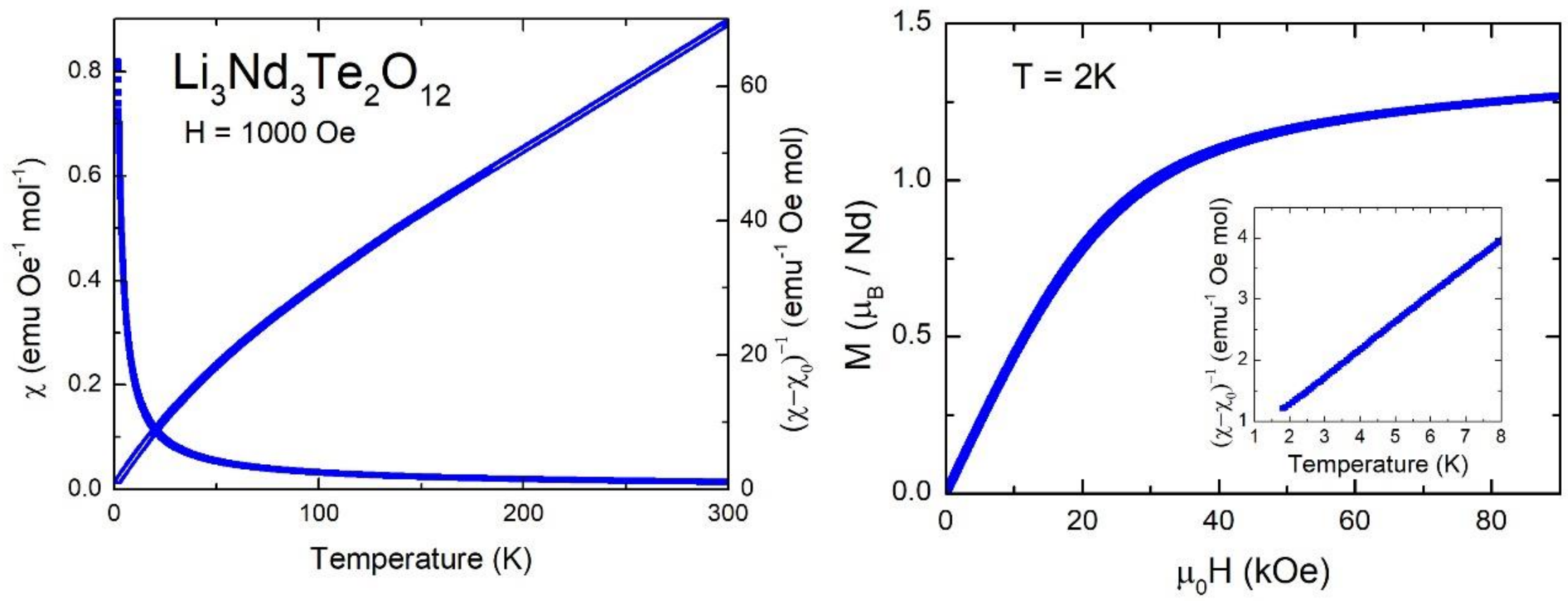

Li3Nd3Te2O12
H = 1000 Oe
χ (emu Oe-1 mol-1)
(χ−χ0)−1 (emu-1 Oe mol)
Temperature (K)
T = 2K
M (μB / Nd)
μ0H (kOe)
(χ−χ0)−1 (emu-1 Oe mol)
Temperature (K)

**Figure 6.**

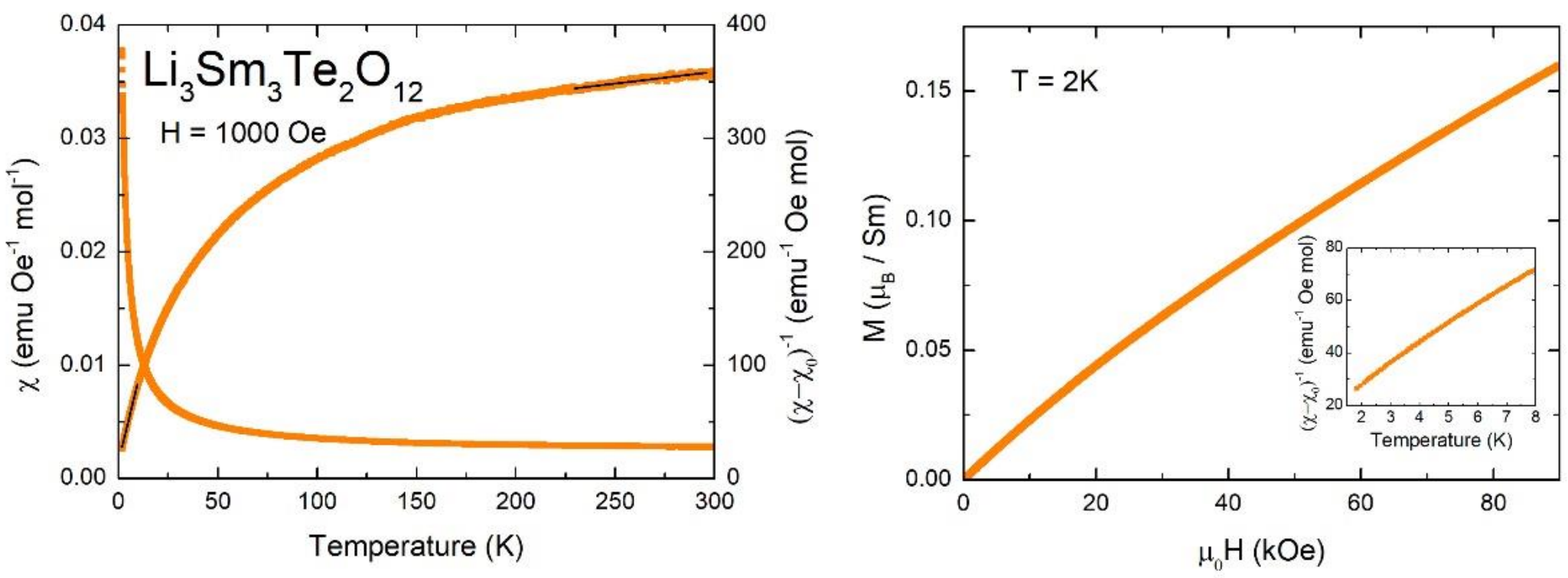
Li3Sm3Te2O12
H = 1000 Oe
χ (emu Oe-1 mol-1)
(χ−χ0)-1 (emu-1 Oe mol)
Temperature (K)
T = 2K
M (μB / Sm)
μ0H (kOe)
(χ−χ0)-1 (emu-1 Oe mol)
Temperature (K)

**Figure 7.**

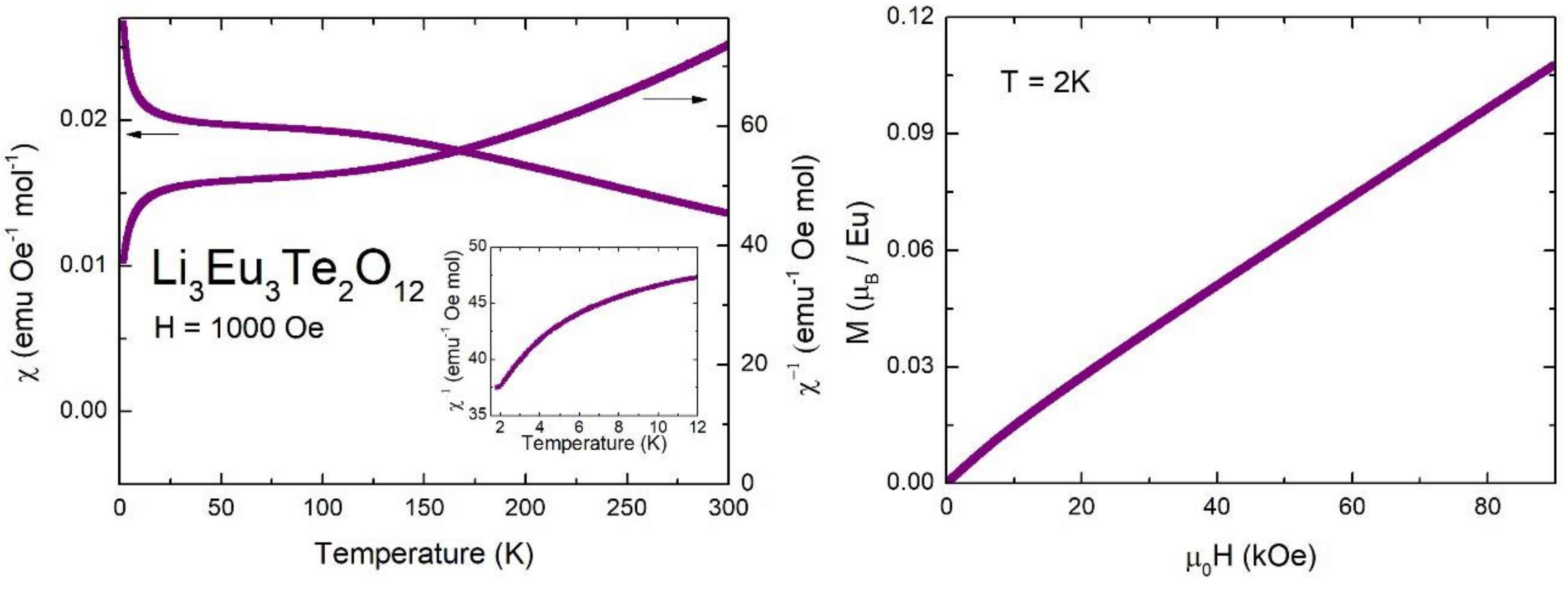

Li3Eu3Te2O12
H = 1000 Oe
χ (emu Oe-1 mol-1)
χ-1 (emu-1 Oe mol)
Temperature (K)
T = 2K
M (μB / Eu)
μ0H (kOe)

**Figure 8.**

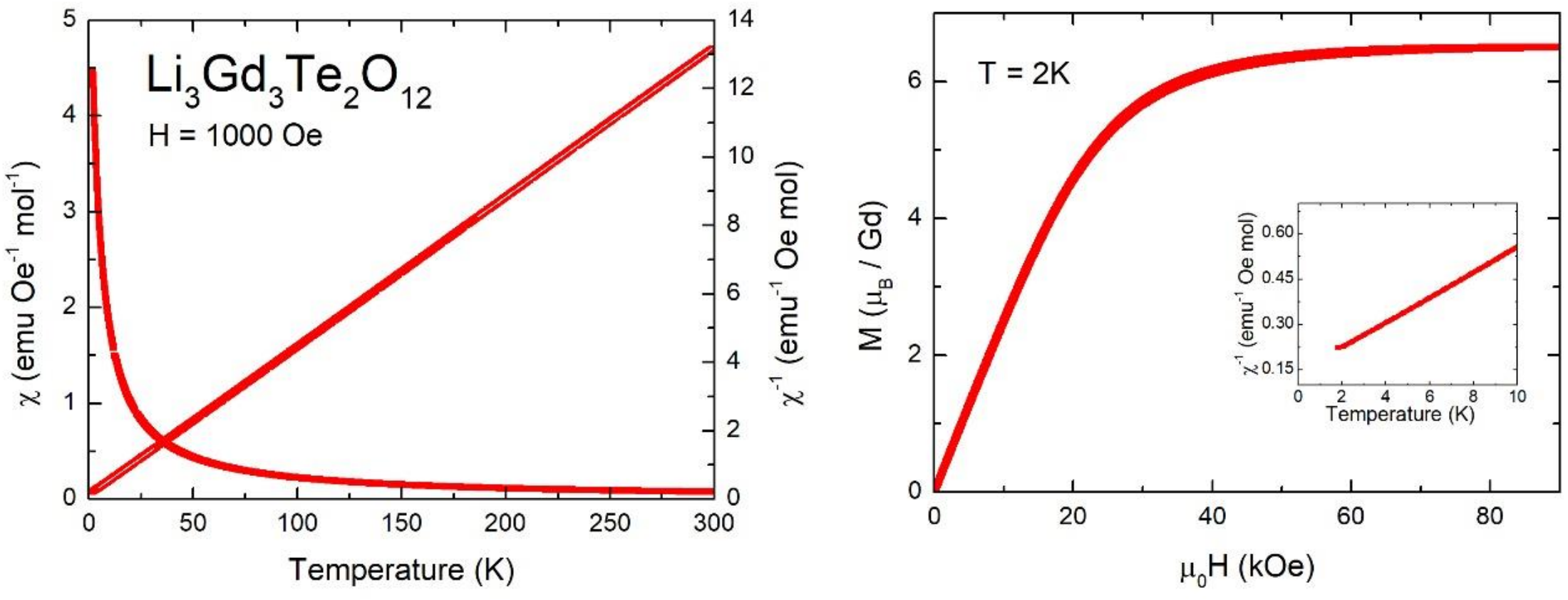

Li3Gd3Te2O12
H = 1000 Oe
χ (emu Oe-1 mol-1)
χ-1 (emu-1 Oe mol)
Temperature (K)
T = 2K
M (μB / Gd)
μ0H (kOe)
χ-1 (emu-1 Oe mol)
Temperature (K)

**Figure 9.**

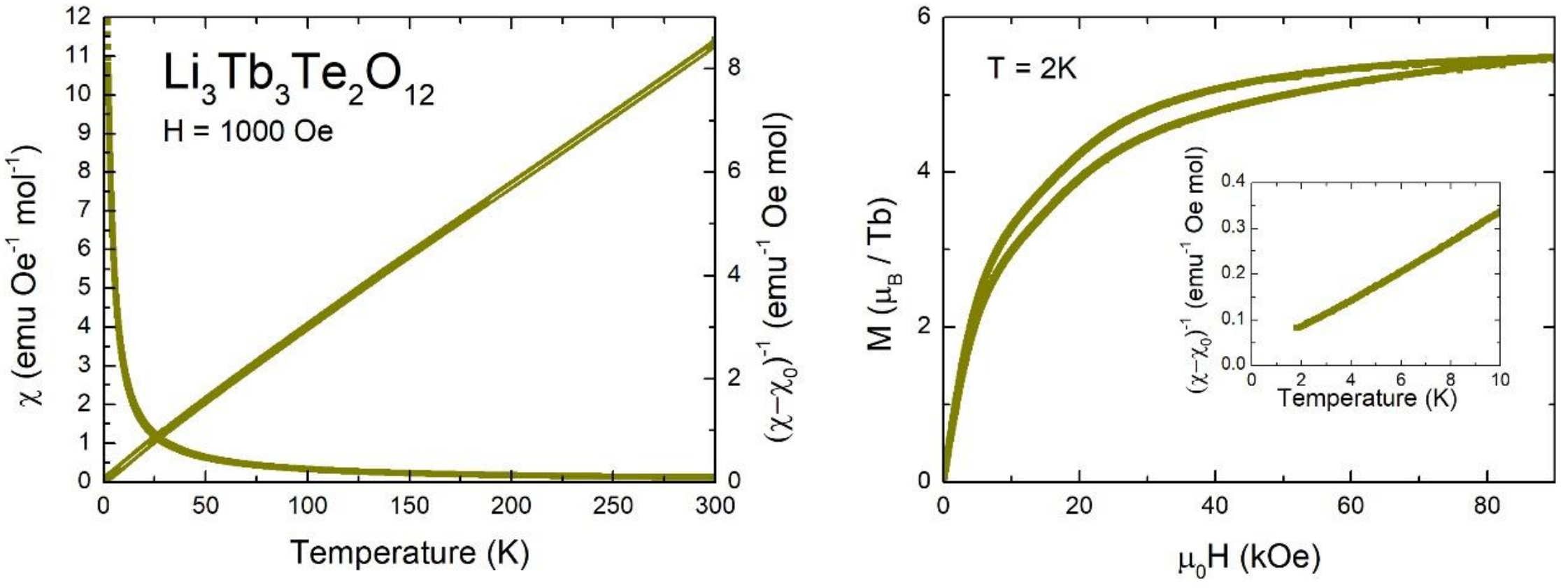
Li3Tb3Te2O12
H = 1000 Oe
χ (emu Oe-1 mol-1)
(χ−χ0)-1 (emu-1 Oe mol)
Temperature (K)
T = 2K
M (μB / Tb)
μ0H (kOe)
(χ−χ0)-1 (emu-1 Oe mol)
Temperature (K)

**Figure 10.**

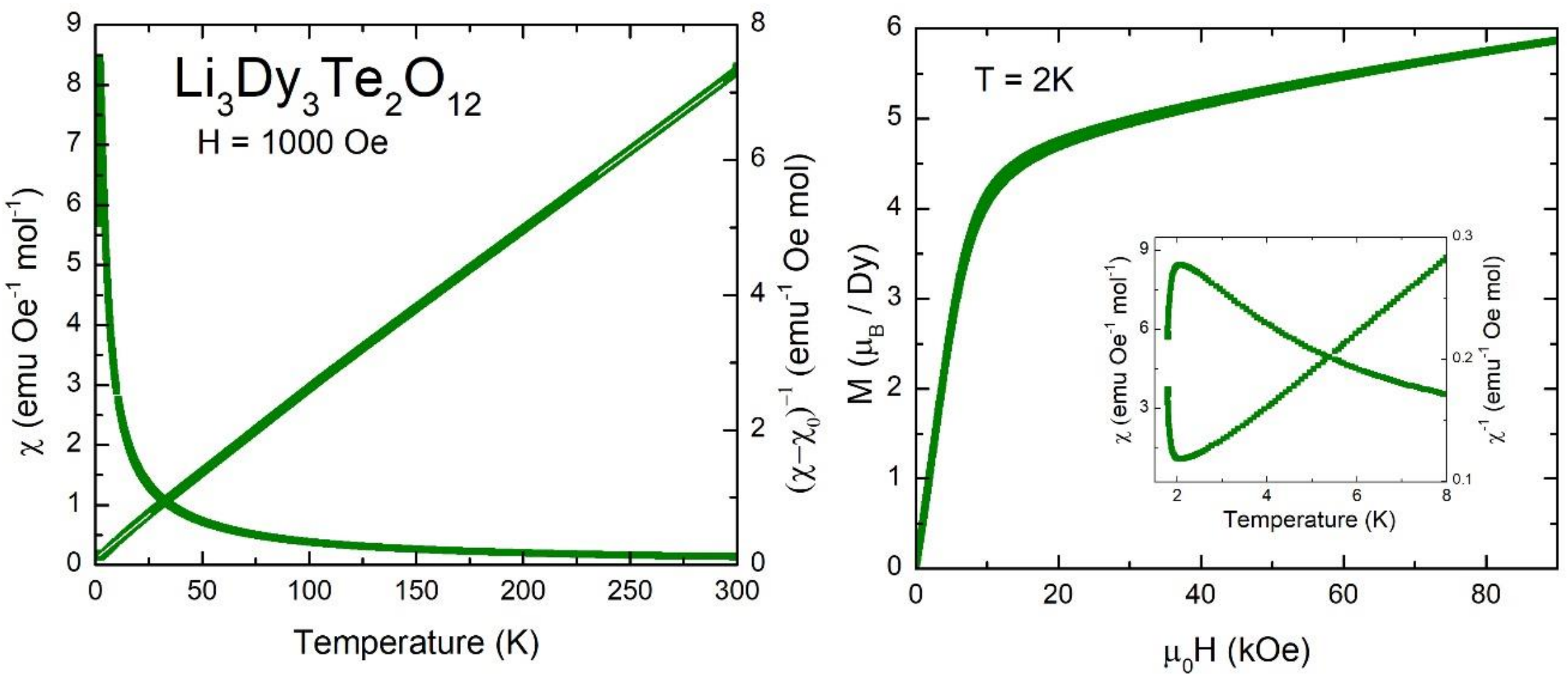

$Li_3Dy_3Te_2O_{12}$
H = 1000 Oe
χ (emu Oe⁻¹ mol⁻¹)
(χ−χ0)⁻¹ (emu⁻¹ Oe mol)
Temperature (K)
T = 2K
M (μB / Dy)
μ0H (kOe)
χ (emu Oe⁻¹ mol⁻¹)
χ⁻¹ (emu⁻¹ Oe mol)
Temperature (K)

**Figure 11.**

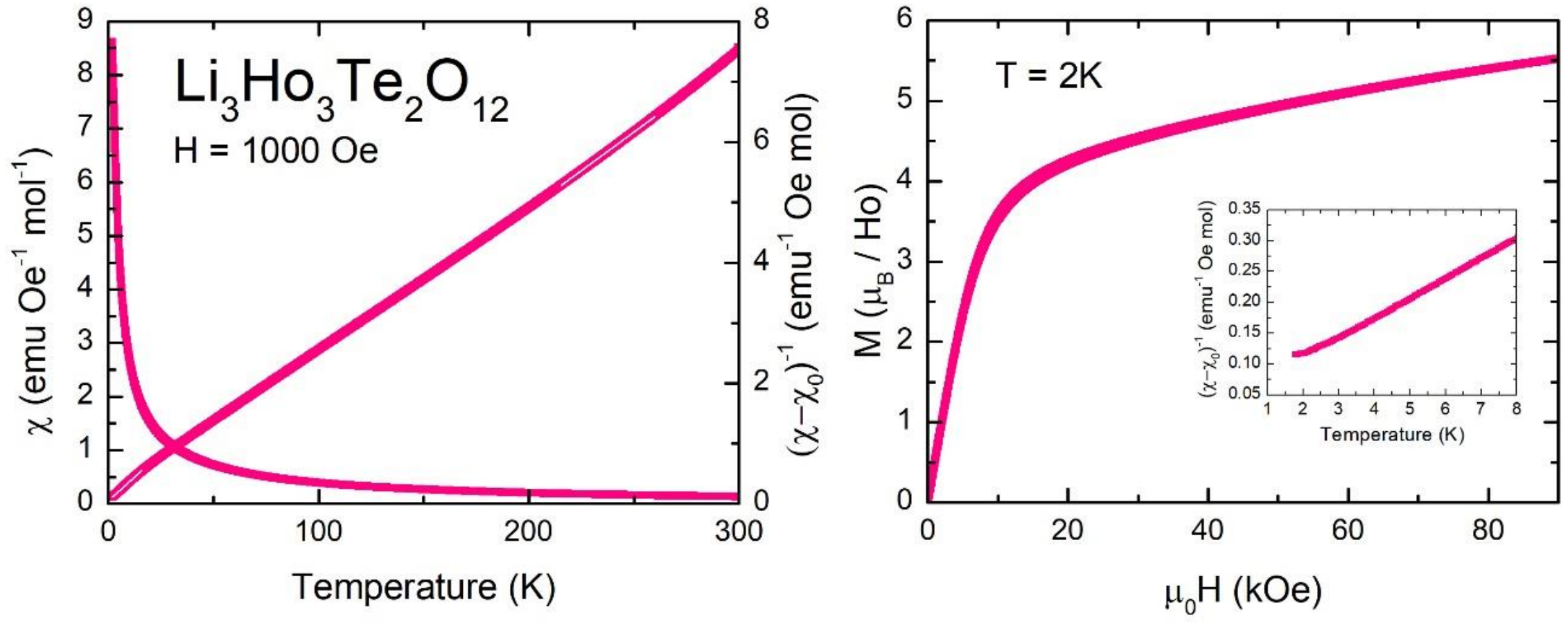

Li3Ho3Te2O12
H = 1000 Oe
χ (emu Oe-1 mol-1)
(χ−χ0)-1 (emu-1 Oe mol)
Temperature (K)
T = 2K
M (μB / Ho)
μ0H (kOe)
(χ−χ0)-1 (emu-1 Oe mol)
Temperature (K)

**Figure 12.**

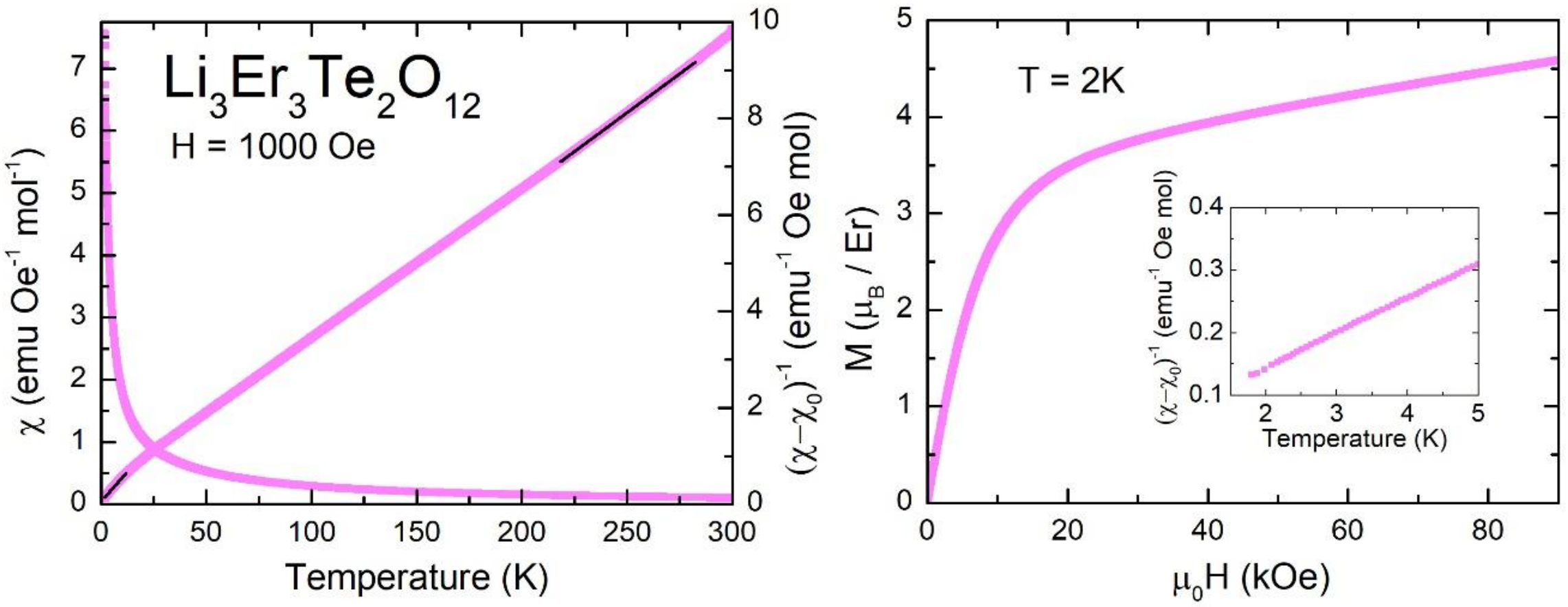

Li3Er3Te2O12
H = 1000 Oe
χ (emu Oe-1 mol-1)
(χ−χ0)-1 (emu-1 Oe mol)
Temperature (K)
T = 2K
M (μB / Er)
μ0H (kOe)
(χ−χ0)-1 (emu-1 Oe mol)
Temperature (K)

**Figure 13.**

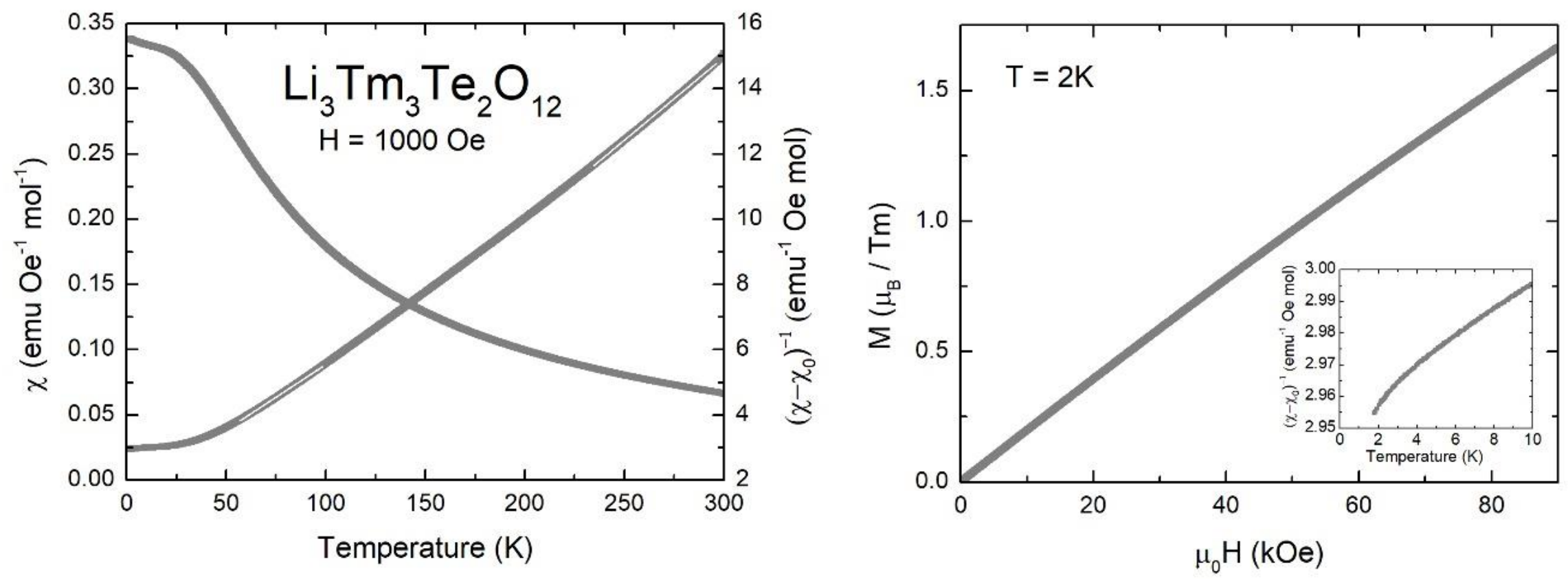
Li3Tm3Te2O12
H = 1000 Oe
χ (emu Oe-1 mol-1)
(χ−χ0)-1 (emu-1 Oe mol)
Temperature (K)
T = 2K
M (μB / Tm)
μ0H (kOe)
(χ−χ0)-1 (emu-1 Oe mol)
Temperature (K)

**Figure 14.**

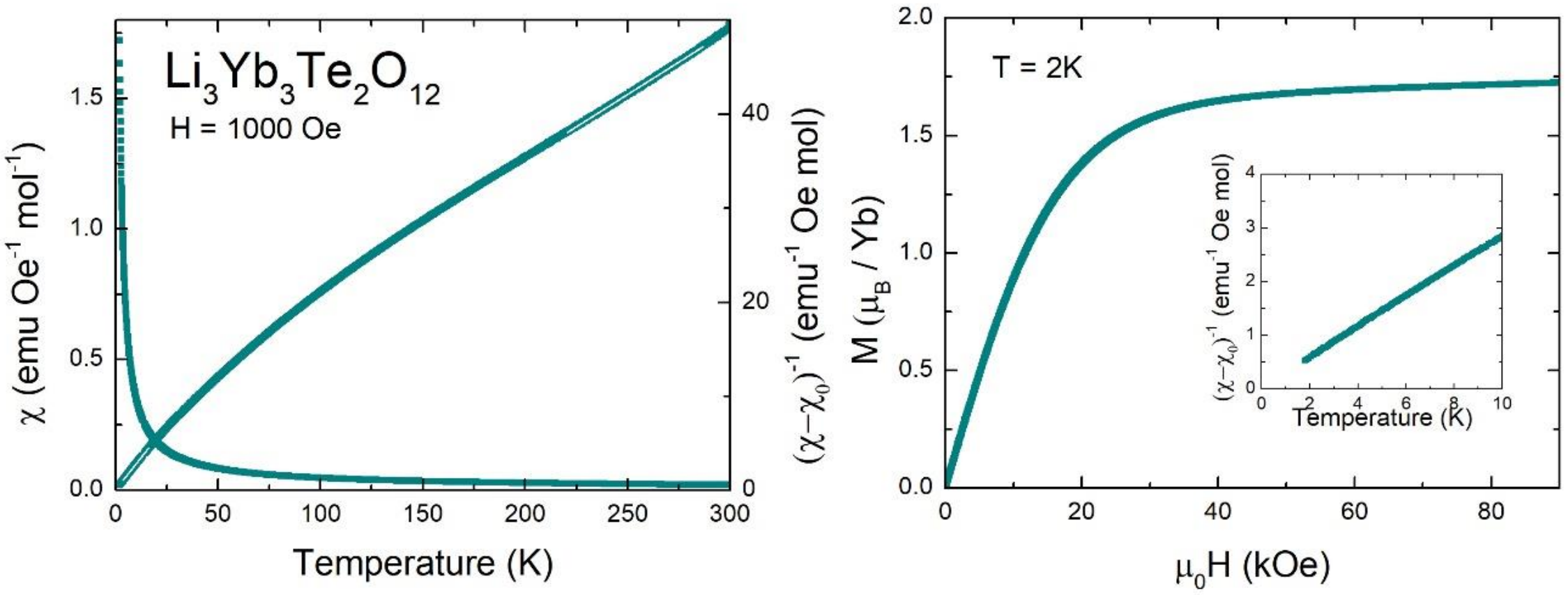

$Li_3Yb_3Te_2O_{12}$
H = 1000 Oe
χ (emu Oe$^{-1}$ mol$^{-1}$)
$(\chi-\chi_0)^{-1}$ (emu$^{-1}$ Oe mol)
Temperature (K)
T = 2K
M ($\mu_B$ / Yb)
$\mu_0$H (kOe)

**Figure 15.**

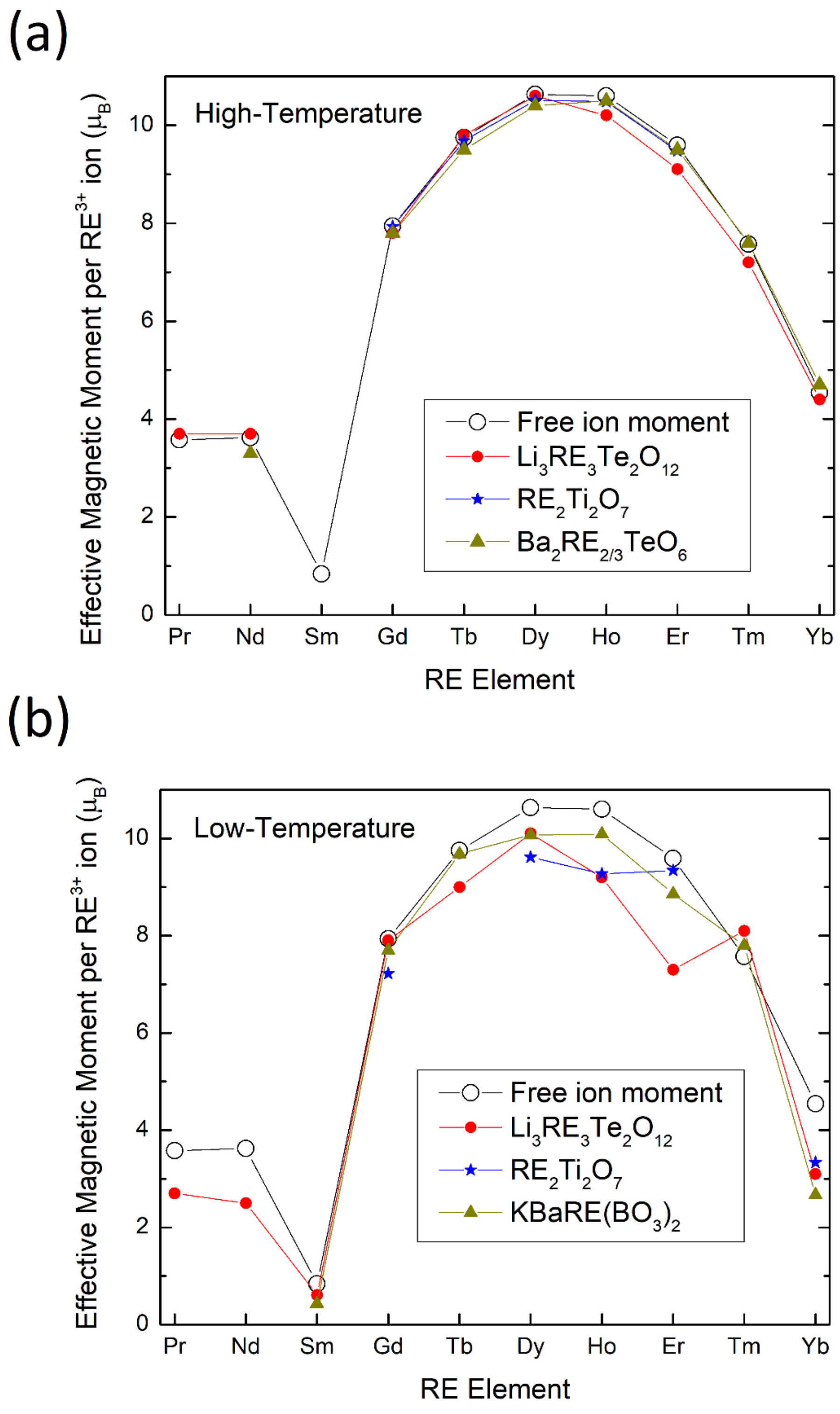

**Figure 16**

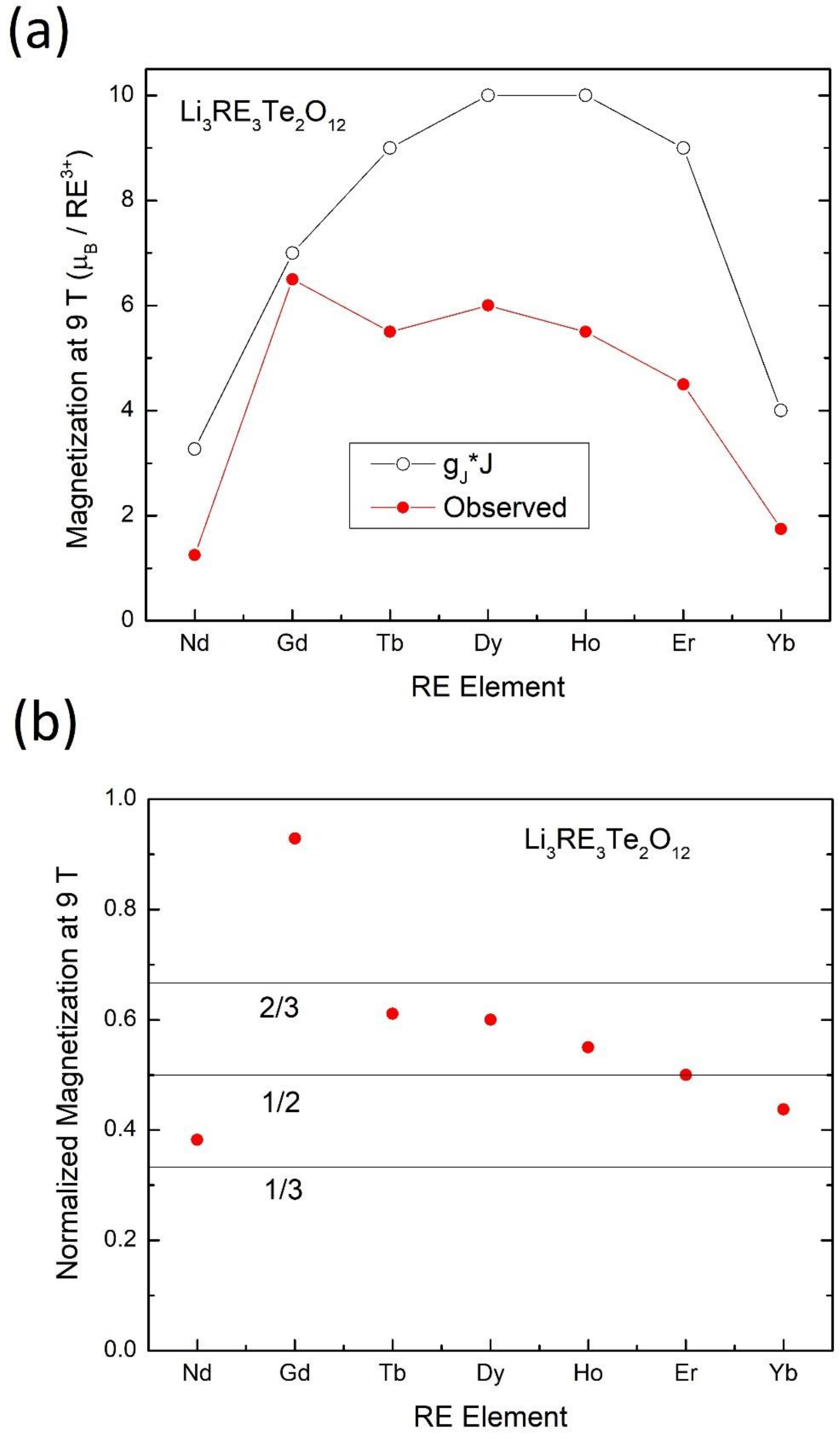

**Figure 17**